\documentclass
[superscriptaddress,secnumarabic,amssymb,amsmath,nobibnotes,aps,prd,showkeys,showpacs,nofootinbib,onecolumn,noshowpacs]{revtex4}%
\usepackage{graphicx}
\usepackage{subfigure}
\usepackage{epstopdf}
\usepackage{epsf}
\usepackage{bm}
\usepackage{amsmath}
\usepackage{amsfonts}
\usepackage{amssymb}
\usepackage[T1]{fontenc}
\usepackage[colorlinks=true,             linkcolor=blue, urlcolor=blue,
citecolor=blue]{hyperref}
\usepackage[colorinlistoftodos]{todonotes}%
\providecommand{\U}[1]{\protect\rule{.1in}{.1in}}

\newcommand{\be}{\begin{equation}}
\newcommand{\ee}{\end{equation}}

\newcommand{\mincir}{\raise
-3.truept\hbox{\rlap{\hbox{$\sim$}}\raise4.truept\hbox{$<$}\ }}
\newcommand{\magcir}{\raise
-3.truept\hbox{\rlap{\hbox{$\sim$}}\raise4.truept\hbox{$>$}\ }}

\begin{document}
\title{Exact black hole solutions in Einstein-Aether Scalar field theory}
\author{N. Dimakis}
\email{nsdimakis@scu.edu.cn ; nsdimakis@gmail.com}
\affiliation{Center for Theoretical Physics, College of Physics, Sichuan University, Chengdu 610064, China}
\author{Genly Leon}
\email{genly.leon@ucn.cl}
\affiliation{Departamento de Matem\'{a}ticas, Universidad Cat\'{o}lica del Norte, Avda.
Angamos 0610, Casilla 1280 Antofagasta, Chile.}
\author{Andronikos Paliathanasis}
\email{anpaliat@phys.uoa.gr}
\affiliation{Institute of Systems Science, Durban University of Technology, Durban 4000,
South Africa}
\affiliation{Instituto de Ciencias F\'{\i}sicas y Matem\'{a}ticas, Universidad Austral de
Chile, Valdivia 5090000, Chile}

\begin{abstract}
We present exact solutions in Einstein-aether theory in a static spherically
symmetric background space with a spacelike aether field, as a difference with the usual  selection of timelike aether field. We assume a coupling
between the scalar field and the aether field introduced in the aether
coefficients. The exact spacetimes describe hairy  black hole solutions for which the
limits of the Schwarzschild, de-Sitter Schwarzschild and Reissner-Nordstr\"{o}m
metrics are recovered.

\end{abstract}
\date{\today}
\maketitle

\section{Introduction}

\label{sec1}

Einstein-Aether gravitational theory is a modification of Einstein's General
Relativity where the kinematic quantities of a unitary time-like vector field,
known as aether, is introduced in the gravitational Action Integral
\cite{jacob,DJ2,Carru,carroll,mm1,hh1,hh2,hh3}. The introduction of the aether
field indicates the selection of a preferred frame which means that there is a
violation of the Lorentz symmetry \cite{ea1}. Specifically, quadratic
quantities of the kinematic terms of the aether field are introduced involving
no more than two derivatives which lead to a second-order theory as the case
of General Relativity. Another important characteristic of the Einstein-Aether
theory is that it describes the classical limit of Ho\v{r}ava gravity
\cite{Horava:2009uw,esf,esf1}. More precisely, every hypersurface-orthogonal
Einstein-Aether solution is a Ho\v{r}ava solution \cite{TJab13}. The
equivalence between the Einstein-Aether and Ho\v{r}ava theories is true in
terms of exact solution and only. In what regards however other generic results, which follow from the direct form of the field
equations, this is not the case \cite{sot01}.

In order to study the effects of Lorentz violation in scalar field theories,
it has been proposed the introduction of a scalar field in Einstein-Aether
action. The most general gravitational Action Integral with an arbitrary
coupling between the inflaton scalar field and the aether field is given in
\cite{DJ}. A specific form of this Action Integral was proposed by Kanno and Soda
in \cite{Kanno:2006ty} where the scalar field is introduced as a quintessence
and the couplings of the aether with the gravitational field
are functions of the scalar field. This specific model was put forth in order to
study the impact of the Lorentz violation on the inflationary scenario. Indeed, it
was found that in this model the inflationary stage is divided into two parts;
the Lorentz violating stage and the standard slow-roll stage. In the Lorentz
violating stage the universe expands as an exact de Sitter spacetime, although
the inflaton field is rolling down the potential. Cosmological studies on
isotropic and anisotropic spacetimes for the Einstein-Aether theory can be
found in
\cite{ea2,Barrow:2012qy,Carruthers:2010ii,Latta:2016jix,Alhulaimi:2017ocb,VanDenHoogen:2018anx,Roumeliotis:2019tvu,Roumeliotis:2018ook,Paliathanasis:2020bgs,Paliathanasis:2019pcl,palgn1,palgn2}
and references therein.

As far as compact objects are concerned, the dynamics of the field equations
for inhomogeneous spherically symmetric models in Einstein-Aether theory are
studied in \cite{Coley:2015qqa} for a non-comoving perfect fluid source.
Spherically symmetric spacetimes with a perfect fluid and a scalar field are
investigated in \cite{Coley:2019tyx} and new exact solutions are presented. The
integrability of the field equations for static spherical symmetric spacetimes
in Einstein-Aether theory with a perfect fluid is investigated in
\cite{Leon:2019jnu} in addition to applying the modified Tolman--Oppenheimer--Volkoff approach.

In the case of vacuum, exact black holes solutions in Einstein-Aether theory
are determined in \cite{bhea1}, where it is shown that the theory possesses
spin-0, spin-1 and spin-2 \ metric modes whose speeds depend on the four coupling
coefficients of the aether field. These
solutions have similarities with the Schwarzschild spacetime outside the
horizon for a wide range of couplings. Black hole solutions with parameterized post-Newtonian (PPN)
parameters identical to those of General Relativity are presented in \cite{bhea2}. In
general, Einstein-Aether black holes provide different evolution of
gravitational perturbations from that of a Schwarzschild black hole
\cite{bhea3}. Charged black hole solutions in an $n$-dimensional
spacetime with or without the cosmological constant term were recently investigated in
\cite{bhea4}. The quasi-normal models for Einstein-Aether black hole
solutions are studied numerically in \cite{bhea5} and \cite{bhea6},
while the matter accretion in Einstein-Aether black hole solutions is analyzed in
\cite{bhea7}. Some rotating spherically symmetric spacetimes in
Einstein-Aether theory are given in \cite{bhea8}. For other studies on
compact stars in Einstein-Aether gravity we refer the reader to
\cite{bh01,bh02,bh03,bh04,bh05,bh06} and references therein.

In this work we investigate the existence of exact solutions in
Einstein-Aether scalar field theory for a static spherically symmetric
background space. For the gravitational action we consider the model proposed
in \cite{Kanno:2006ty}. We require the field equations to admit a point-like
Lagrangian which indicates that the aether field should be space-like. Models with a space-like aether field have been previously studied in the context of a small violation of the rotation invariance in the early universe \cite{Carrollsp,Dulaney} or in theories in higher dimensions \cite{nn5,nn4,nn3}. The
gravitational field equations depend on three unknown functions of the scalar
field, two are the coupling functions between the aether and the scalar
fields, while the third function is the scalar field potential. For certain choices of these functions the system possesses enough integrals of motion for the exact solution to be derived. We show that under specific relations among the constants of integration black hole solutions emerge. What is more, under taking specific limiting values of the parameters, one is led to the known solutions of the Einstein(-Maxwell) equations of General Relativity.

The outline of the paper is as follows: In Section \ref{sec2} the Action Integral and the general setting of the gravitational theory of our
consideration is presented. Section \ref{sec3} includes the main result of
this work; we present the static, spherically symmetric spacetime which consists the general exact solution of the field equations and we
give the conditions under which this solution describes black holes of the Einstein-Aether
scalar field theory. In Section \ref{sec4} we prove the existence of circular
orbits for massive particles. Finally, in Section \ref{sec5} we summarize our
results and we draw our final conclusions.

\section{Einstein-Aether Scalar field theory}

\label{sec2}

For the gravitational Action Integral we consider the Einstein-Aether scalar
field theory
\begin{equation}
S=\int dx^{4}\sqrt{-g}\left(  \frac{\mathcal{R}}{2}-\frac{1}{2}g^{\mu\nu}\nabla_{\mu}\phi%
\nabla_{\nu}\phi-V\left(  \phi\right) + \mathcal{L}_{Aether} \right) ,\label{ac.01}%
\end{equation}
where $\mathcal{R}$ is the Ricci scalar, $g=\mathrm{Det}(g_{\mu\nu})$ the metric determinant, $V(\phi)$ the potential of the scalar field and $\mathcal{L}_{Aether}$ indicates the  Aether field Lagrangian density
\cite{Kanno:2006ty}
\begin{equation}
  \mathcal{L}_{Aether}  =  K^{\alpha\beta}_{\;\;\mu\nu} \nabla_\alpha u^\mu \nabla_\beta u^\nu + \lambda_0 (u^\mu u_\mu + \varepsilon) \label{ac.02}%
\end{equation}
where
\begin{equation}\label{Ktensor}
  K^{\alpha\beta}_{\;\;\mu\nu}  = - \left(\beta_1(\phi) g^{\alpha\beta} g_{\mu\nu} +\beta_2(\phi) \delta^\alpha_{\; \mu}\delta^\beta_{\; \nu} + \beta_3(\phi) \delta^\alpha_{\; \nu}\delta^\beta_{\; \mu} + \beta_4(\phi) u^\alpha u^\beta g_{\mu\nu}\right)  .
\end{equation}

The function $\lambda_0$ is a Lagrange multiplier and $\varepsilon=\pm1,0$ the constant which
serves to fix the measure of the velocity of the aether field as $u^{\mu}u_\mu = -\varepsilon$. On the other hand, the
coupling functions $\beta_{1}\left(  \phi\right)  ,~\beta_{2}\left(
\phi\right)  ,~\beta_{3}\left(  \phi\right)  ~$and $\beta_{4}\left(
\phi\right)  $ define the coupling between the aether field and the
gravitational field. In the typical Einstein-Aether theory the Aether field $u^\mu$ is assumed to be time-like
\cite{DJ2,Carru} which means $\varepsilon=1$ in the Lagrangian density \eqref{ac.02}. However, there exist modifications of the theory where
null-like, i.e. $\varepsilon=0~$\cite{nn1,nn2}, or
space-like $\varepsilon=-1$ \cite{nn3,nn4,nn5} fields are considered. Although various generic models that include space-like vector fields are known to be unstable \cite{unstable1}, it is claimed that under certain conditions stable configurations can be constructed \cite{stable1}. A space-like aether field is what we consider in this work by the means to extract from the action \eqref{ac.01} a valid point-like Lagrangian, hence from now on we assume $\varepsilon=-1$.

By defining $J^\mu_{\;\alpha} = K^{\mu\nu}_{\;\;\alpha\beta} \nabla_\nu u^\beta$  \cite{carroll}, the field equations for the metric can be written as:
\begin{equation} \label{EinA}
  \mathcal{R}_{\mu\nu} - \frac{1}{2} \mathcal{R} g_{\mu\nu} = T_{\mu\nu}^{Aether} + T^{Scalar}_{\mu\nu}
\end{equation}
with
\begin{equation}
  \begin{split}
    T_{\mu\nu}^{Aether} = & 2 \beta_1 \left( \nabla_\mu u^\alpha \nabla_\nu u_\alpha - \nabla^\alpha u_\mu \nabla_\alpha u_\nu \right) - 2 \left[\nabla_\alpha\left(u_{(\mu}J^\alpha_{\;\nu)}\right)+ \nabla_\alpha \left(u^\alpha J_{(\mu\nu)}\right) - \nabla_\alpha \left( u_{(\mu} J_{\nu)}^{\; \alpha}\right) \right] \\
    & -2\beta_4 u^\alpha u^\beta \nabla_\alpha u_\mu \nabla_\beta u_\nu + g_{\mu\nu}\mathcal{L}_{Aether} + 2 \left[u_\beta \nabla_\alpha J^{\alpha\beta} +  \beta_4 u^\alpha u^\beta \nabla_\alpha u_\kappa  \nabla_\beta u^\kappa  \right]u_\mu u_\nu
  \end{split}
\end{equation}
and
\begin{equation}
  T^{Scalar}_{\mu\nu} = \nabla_\mu \phi \nabla_\nu \phi - \frac{1}{2} g_{\mu\nu} \left( \nabla_\alpha \phi \nabla^\alpha \phi + 2V(\phi) \right)
\end{equation}
the energy momentum tensors for the aether and scalar fields respectively. Additionally there also exist the equation of motion for the scalar $\phi$, which is
\begin{equation} \label{scfeq}
  \nabla_\mu\nabla^\mu \phi - \frac{d V}{d\phi} - \frac{d \beta_1}{d\phi} \nabla^\nu u^\mu \nabla_\nu u_\mu - \frac{d \beta_2}{d\phi} \left(\nabla_\mu u^\mu\right)^2 - \frac{d \beta_3}{d\phi} \nabla^\mu u^\nu \nabla_\nu u_\mu - \frac{d \beta_4}{d\phi} u^\alpha u^\beta \nabla_\alpha u^\mu \nabla_\beta u_\mu =0,
\end{equation}
and for the aether field $u^\mu$ that leads to
\begin{equation} \label{lamfeq}
  \nabla_\mu J^{\mu\nu} + \beta_4 u^\kappa \nabla_\kappa u_\lambda \nabla^\nu u^\lambda = \lambda_0 u^\nu.
\end{equation}
When the aether field has the additional property $u^\alpha \nabla_\alpha u^\mu =0$, then the fourth function $\beta_4(\phi)$ becomes irrelevant since all its contributions are trivial. Under this condition it can thus be eliminated from the above equations (see the corresponding relations in \cite{carroll}).

In cosmological studies the Action Integral (\ref{ac.01}), has the property
that the gravitational field equations admit a point-like Lagrangian. The
latter property is extremely useful in the application of well known results and techniques from
classical mechanics regarding Noether's theorem. Solutions of this type regarding a
Friedmann--Lema\^{\i}tre--Robertson--Walker and a Bianchi I
spacetime are presented in \cite{palgn1,palgn2}.

\subsection{Static spherically symmetric spacetime}

In this work, for the background space we consider the static, spherically
symmetric spacetime with line element%
\begin{equation} \label{genlineel}
ds^{2}=-e^{2(\beta(r)+\lambda(r))}dt^{2}+N(r)dr^{2}+e^{2\lambda(r)-\beta
(r)}\left(  d\theta^{2}+\sin^{2}\theta d\varphi^{2}\right)
\end{equation}
and the space-like velocity for the aether
\begin{equation}
u_{\mu}=(0,N(r),0,0),\label{velsph}%
\end{equation}
for which additionally it holds that $u^\alpha \nabla_\alpha u^\mu =0$. Hence the function $\beta_4(\phi)$ is excluded from our analysis.

With the above conditions we derive the point-like Lagrangian
\begin{equation}
L\left(  N,\beta,\beta^{\prime},\lambda,\lambda^{\prime}\right)
= \frac{e^{3\lambda}}{2N} \left[-6F(\phi)\lambda^{\prime2} + \frac{3%
M(\phi)\beta^{\prime2}}{4}- \phi^{\prime2}  \right]%
+N \left(e^{\beta+\lambda}-e^{3\lambda}V(\phi) \right),\label{Lag}%
\end{equation}
where the prime denotes total derivative with respect to the variable $r$. The
functions $F\left(  \phi\right)  ,$ $M\left(  \phi\right)  $ are defined as
\begin{equation}
F\left(  \phi\right)  = \beta_{1}\left(  \phi\right)  +3\beta
_{2}\left(  \phi\right)  +\beta_{3}\left( \phi\right) -1
,\label{ac.05}%
\end{equation}%
\begin{equation}
M\left(  \phi\right)  =-2\left[  1 + 2\left(  \beta_{1}\left(  \phi\right)
+\beta_{3}\left(  \phi\right)  \right)  \right] . \label{ac.06}%
\end{equation}
The gravitational field equations \eqref{EinA} are equivalent to the Euler-Lagrange
equations~$\frac{d}{dr}\left(  \frac{\partial L}{\partial\beta^{\prime}%
}\right)  -\frac{\partial L}{\partial\beta}=0$,$~\frac{d}{dr}\left(
\frac{\partial L}{\partial\lambda^{\prime}}\right)  -\frac{\partial
L}{\partial\lambda}=0$, with constraint equation $\frac{\partial L}{\partial
N}=0$. For the scalar field $\phi$ we assume that it
inherits the Killing symmetries of the background space, which means that the
equation of motion \eqref{scfeq} is given by the Euler-Lagrange
equation $\frac{d}{dr}\left(  \frac{\partial L}{\partial\phi^{\prime}}\right)
-\frac{\partial L}{\partial\phi}=0$. Last but not least, the field equation \eqref{lamfeq} just serves to define the multiplier $\lambda_0$. Having performed this consistency check we can concentrate our analysis on the reduced system described by \eqref{Lag}.

\section{Black hole solutions}

\label{sec3}

The nonlinear gravitational field equations depend on three unknown functions,
namely the coupling functions $F\left(  \phi\right)  ,~M\left(  \phi\right)  $
and the scalar field potential $V\left(  \phi\right)  $. In the following we
consider specific functional forms of these unknown functions so as to
extract closed-form solutions for the field equations.

\subsection{The generic $F(\phi)=\mu\phi^{2}$, $M(\phi)=\nu\phi^{2}$,
$V(\phi)=0$ case}

In the particular case where $F(\phi)=\mu\phi^{2}$, $M(\phi)=\nu\phi^{2}$ and $V(\phi)=0$, with $\mu,\nu$ constants, the system admits three linear in the velocities integrals of motion:
\begin{align}
  I_1 = & \frac{e^{3 \lambda} \phi^2}{N} \left( 2\mu \lambda^{\prime} + \nu \beta^{\prime}\right), \\
  I_2 = & \frac{e^{3 \lambda} \phi}{N} \left(3\mu \, \phi \lambda^{\prime} - \phi^{\prime} \right) , \\
  I_3 = & \frac{e^{3 \lambda} \phi }{N} \left[ \phi  \left( \left(3 \mu  \nu  \lambda-\nu  \ln \phi\right)\beta^{\prime} - \left(3 \mu  \nu  \beta+2 \mu  \ln \phi\right)\lambda^\prime \right)+ \left(\nu  \beta+2 \mu  \lambda\right)\phi^\prime \right] .
\end{align}
The fact that a quadratic dependence of $F$ and $M$ in $\phi$ gives rise to this type of conservation laws is also known from the cosmological case, see the recent \cite{GenlyAndrNik}.

The above conservation laws can be used in conjunction to the field equations in order to derive the general solution of the system. We avoid the presentation of the cumbersome but straightforward procedure and we focus in the end result which leads to the line element:
\begin{equation}%
\begin{split}
ds^{2}= &  -\frac{e^{A_{1}r}}{\left[  \cosh\left(  \sqrt{\kappa_{1}}r\right)
\right]  ^{\frac{2(2\mu-\nu)}{\mu(1-3\nu)-2\nu}}}dt^{2}+\frac{\nu^{2}%
e^{\kappa_{3}-A_{1}r}}{\left[  \cosh\left(  \sqrt{\kappa_{1}}r\right)
\right]  ^{\frac{6\nu(2\mu+1)}{\mu(3\nu-1)+2\nu}}}dr^{2}\\
&  +\frac{\kappa_4 e^{A_{2}r}}{\left[  \cosh\left(  \sqrt{\kappa_{1}}r\right)
\right]  ^{\frac{2(\mu+\nu)}{\mu(3\nu-1)+2\nu}}}\left(  d\theta^{2}+\sin
^{2}\theta d\varphi^{2}\right)
\end{split}
\label{lineelgen}%
\end{equation}
and the scalar field
\begin{equation}
\phi(r)=\frac{e^{A_{3}r+\frac{\kappa_{3}}{2}}  \sqrt{2\nu}\sqrt{\mu(3\nu-1)+2\nu}  }{\sqrt{3\kappa_{1}\kappa_4 \mu}\left[  \cosh\left(  \sqrt{\kappa_{1}}r\right)  \right]^{\frac{3\mu\nu}{\mu(1-3\nu)-2\nu}}},
\end{equation}
where $\kappa_4,\kappa_{i},A_{i}$, $i=1,2,3$ are all constants. The latter three
constants, $A_{1}$, $A_{2}$, $A_{3}$, are not independent, but are the
combinations:
\begin{subequations}
\label{constantA1}%
\begin{align}
A_{1} &  =\frac{2\kappa_{2}\nu}{3\mu+2}+\frac{2\sqrt{3\nu}(2\mu+1)%
\sqrt{2\kappa_{2}^{2}\nu(3\mu\nu-\mu+2\nu)+3\kappa_{1}\mu(3\mu+2)}}%
{(3\mu+2)(\mu(3\nu-1)+2\nu)} , \\
A_{2} &  =\frac{2\kappa_{2}\nu}{3\mu+2}-\frac{2\sqrt{3\nu}(\mu+1)%
\sqrt{2\kappa_{2}^{2}\nu(\mu(3\nu-1)+2\nu)+3\kappa_{1}\mu(3\mu+2)}}%
{(3\mu+2)(\mu(3\nu-1)+2\nu)} , \\
A_{3} &  =-\frac{2\kappa_{2}\nu}{3\mu+2}-\frac{\sqrt{3\nu}\mu%
\sqrt{2\kappa_{2}^{2}\nu(3\mu\nu-\mu+2\nu)+3\kappa_{1}\mu(3\mu+2)}}%
{(3\mu+2)(\mu(3\nu-1)+2\nu)} .%
\end{align}

In order to judge if we have a black hole solution we first need to make a
transformation of the form
\end{subequations}
\begin{equation}
\frac{\kappa_4 e^{A_{2}r}}{\left[  \cosh\left(  \sqrt{\kappa_{1}}r\right)
\right]  ^{\frac{2(\mu+\nu)}{\mu(3\nu-1)+2\nu}}}\mapsto\tilde{r}%
^{2}\label{transf}%
\end{equation}
to associate the function multiplying the unit sphere part of the metric, $d\Omega^2 = d\theta^{2}+\sin^{2}\theta
d\varphi^{2}$, with some radial distance. The
mapping \eqref{transf} forms an algebraic equation which cannot be
solved for all the values of the constants as $r=r(\tilde{r})$. However, if we
enforce the restriction
\begin{equation}
A_{2}=-\frac{2(\mu+\nu)}{\mu(3\nu-1)+2\nu}\sqrt{\kappa_{1}},\label{specialcon}%
\end{equation}
then the transformation
\begin{equation}
r(\tilde{r})=\frac{1}{2\sqrt{\kappa_{1}}}\ln\left(  2\kappa_4^{\frac{\mu
(3\nu-1)+2\nu}{2(\mu+\nu)}}\tilde{r}^{\frac{-3\mu\nu+\mu-2\nu}{\mu+\nu}%
}-1\right)
\end{equation}
realizes the mapping \eqref{transf} and the resulting line element reads
\begin{equation}%
\begin{split}
ds^{2}= &  -r^{2-\frac{6\mu}{\mu+\nu}}\left(  1-2\kappa_4^{\frac{\mu
(3\nu-1)+2\nu}{2(\mu+\nu)}}r^{\frac{-3\mu\nu+\mu-2\nu}{\mu+\nu}}\right)
^{\frac{A_{1}}{2\sqrt{\kappa_{1}}}+\frac{\nu-2\mu}{\mu(3\nu-1)+2\nu}}dt^{2}\\
&  +\frac{e^{C}r^{\frac{6\mu\nu}{\mu+\nu}}}{\left(  1-2\kappa_4^{\frac{\mu
(3\nu-1)+2\nu}{2(\mu+\nu)}}r^{\frac{-3\mu\nu+\mu-2\nu}{\mu+\nu}}\right)
^{\frac{A_{1}}{2\sqrt{\kappa_{1}}}+\frac{\nu-2\mu}{\mu(3\nu-1)+2\nu}}}%
dr^{2}+r^{2}\left(  d\theta^{2}+\sin^{2}\theta d\varphi^{2}\right)
\end{split}
\label{metbhgen}%
\end{equation}

For simplicity we drop the tilde over $r$, but it is to be understood that
from now on the $r$ appearing is different than the one we had in
\eqref{lineelgen}. In addition, some constant scalings in the $t$ variable
have been made in order to simplify the line element. The constant $C$ is a
reparametrization of the constant $\kappa_{3}$ given by
\begin{equation}
\kappa_{3}=\ln\left(  (-1)^{\frac{A_{1}\mu(3\nu-1)+2A_{1}\nu-6\sqrt{\kappa
_{1}}(2\mu+1)\nu}{2\sqrt{\kappa_{1}}(\mu(3\nu-1)+2\nu)}}\kappa_4^{\frac{3\mu
\nu+\mu+\nu}{\mu+\nu}}\right)  +C.
\end{equation}

Finally, the corresponding scalar field is
\begin{equation}%
\begin{split}
\phi(r) &  =(-1)^{\frac{A_{1}\mu(3\nu-1)+2A_{1}\nu+A_{3}\mu(6\nu-2)+4A_{3}%
\nu-6\sqrt{\kappa_{1}}(\mu+1)\nu}{4\sqrt{\kappa_{1}}(\mu(3\nu-1)+2\nu)}}%
\frac{\sqrt{\frac{2}{3}}(\mu+\nu)\sqrt{3\mu\nu-\mu+2\nu}}{\sqrt{\mu}\sqrt{\nu
}(\mu(3\nu-1)+2\nu)}e^{\frac{C}{2}}\\
&  \times r^{\frac{3\mu\nu}{\mu+\nu}}\left(  1-2\kappa_4^{\frac{\mu(3\nu
-1)+2\nu}{2(\mu+\nu)}}r^{\frac{-3\mu\nu+\mu-2\nu}{\mu+\nu}}\right)  ^{\frac
{1}{2}\left(  \frac{A_{3}}{\sqrt{\kappa_{1}}}+\frac{3\mu\nu}{\mu(3\nu-1)+2\nu
}\right)  }.
\end{split}
\end{equation}

\subsection{Distinguishing black hole solutions}

In order for the line element \eqref{metbhgen} to be able to describe a black hole space-time we need at least to enforce two further conditions: (i)
demand the power of $r$ in the parenthesis to be negative, i.e.
\begin{equation}
\frac{-3\mu\nu+\mu-2\nu}{\mu+\nu}<0\label{cond1}%
\end{equation}
and (ii) at the same time the power of the first parenthesis to assume the value of an odd positive number
\begin{equation}
\frac{A_{1}}{2\sqrt{\kappa_{1}}}+\frac{\nu-2\mu}{\mu(3\nu-1)+2\nu}=2k+1,\quad
k\in\mathbb{N}.\label{cond2}%
\end{equation}
so that the $g_{tt}$ and $g_{rr}$ components of the metric can interchange signs when
crossing the Killing horizon. Of course we also need to impose $\kappa_4>0$, so that such an
horizon exists at a real distance $r=r_{h}=  2^{\frac{\mu +\nu }{\mu  (3 \nu -1)+2 \nu }} \sqrt{\kappa_4} = (2\kappa)^{\frac{\mu +\nu }{\mu  (3 \nu -1)+2 \nu }}$, where from now on for the simplicity of the line element we define $\kappa=\kappa_4^{\frac{\mu (3\nu-1)+2\nu}{2(\mu+\nu)}}$.

The above two conditions can be satisfied for an infinite combination of $\mu$
and $\nu$ values. To demonstrate this, let us take the case $k=0$ in
\eqref{cond2}, so that
\begin{equation}
\frac{A_{1}}{2\sqrt{\kappa_{1}}}+\frac{\nu-2\mu}{\mu(3\nu-1)+2\nu}=1.
\end{equation}

By using the above relation together with \eqref{constantA1} and the necessary
condition \eqref{specialcon}, which is needed for the special solution
\eqref{metbhgen} to exist, we see that they are all compatible for every
nonzero value of $\mu>-1$ and $\kappa_{1}>0$ if $\nu=2\mu$. If we insert the
latter however in \eqref{cond1} we see that we obtain the additional
restriction $\mu>-\frac{1}{2}$.

As a result we have a black hole of the form
\begin{equation}
ds^{2}=-\left(  1-\frac{2\kappa}{r^{2\mu+1}}\right)
dt^{2}+\frac{e^{C}r^{4\mu}}{1-\frac{2\kappa}{r^{2\mu+1}}%
}dr^{2}+r^{2}\left(  d\theta^{2}+\sin^{2}\theta d\varphi^{2}\right)
\label{bhole1}%
\end{equation}
corresponding to the scalar field
\begin{equation}
\phi(r)=\pm\frac{e^{\frac{C}{2}}r^{2\mu}}{\sqrt{-\mu(2\mu+1)}}%
\label{solphibh1}%
\end{equation}
with the restriction $\mu>-\frac{1}{2}$ (and of course $\mu\neq0$). Notice
that if $\mu>0$ the scalar field $\phi$ is imaginary, which makes its
contribution in the action to be that of a phantom field. On the contrary,
when $-\frac{1}{2}<\mu<0$ we have a solution with a canonical scalar field.

In addition, we observe that apart from the constant $\kappa$, which can be associated with the mass of the black hole, the line element \eqref{bhole1} carries a dependence on the constant $C$ that emerges from the matter content, i.e. the scalar field. We thus deduce that this is a hairy black hole since another constant appears in addition to the mass (in this particular case we have not considered an electromagnetic field or a rotating solution for an additional charge or an angular momentum respectively). Through the use of curvature scalars it is a simple task to indeed verify that both $\kappa$ and $C$ are essential for the geometry, i.e. they can not be absorbed with a diffeomorphism \cite{Papadopoulos}. Take for example the triplet $q=(q_1,q_2,q_3)$ with $q_1=\mathcal{R}$, $q_2 = \nabla_\alpha \mathcal{R} \nabla^\alpha \mathcal{R}$ and the Kretschmann scalar $q_3 = \mathcal{R}_{\alpha\beta\gamma\delta}\mathcal{R}^{\alpha\beta\gamma\delta}$; then the matrix $\mathcal{J}_{ij}=\frac{\partial q_i}{\partial v^j}$ with $v=(r,\kappa,C)$, $i,j=1,2,3$, is invertible. As a result, you can in principle use one of the equations defined by the $q_i$ to solve with respect to $r$ and substitute in the remaining couple, then two algebraically independent relations of the form $f_1(q_1,q_2,q_3,\kappa,C)=0$, $f_2(q_1,q_2,q_3,\kappa,C)=0$ will be formed, involving both $\kappa$ and $C$, and with the property of being invariant under local coordinate transformations. Hence, neither $\kappa$ nor $C$ can be eliminated through such a mapping.

We observe that for $\mu\rightarrow0$ and by assuming $e^C\rightarrow 1$, the spacetime (\ref{bhole1}) takes the form
of the Schwarzschild black hole, thus for small values of $\mu$ and $C=0$ the line
element (\ref{bhole1}) becomes%
\begin{equation}
ds^{2}=ds_{Schwarzschild}^{2}+ \mu  \left[-4 \kappa \frac{\ln r}{r} dt^2 + \frac{4 r (r-3\kappa)\ln r}{(r-2\kappa)^2} dr^2\right] + \mathcal{O}(\mu^2)
\end{equation}
Alternatively, we may introduce the transformation $r=(2 \mu +1)^{\frac{1}{2 \mu +1}} R^{\frac{1}{2 \mu +1}}$ to write the line element \eqref{bhole1} as
\begin{equation} \label{bhole1b}
  ds^2 = -\left(1- \frac{2\kappa}{(1+2\mu)R} \right) dt^2 + e^C\left(1- \frac{2\kappa}{(1+2\mu)R} \right)^{-1} dR^2 + (1+ 2 \mu)^{\frac{2}{2 \mu +1}} R ^{\frac{2}{2 \mu +1}}  \left(  d\theta^{2}+\sin^{2}\theta d\varphi^{2}\right).
\end{equation}
The above metric, although it is distinct, it resembles in some parts solutions expressing non-asymptotically flat black holes in the context of Einstein-Maxwell-dilaton gravity presented in \cite{EinMaxdil}.

It is interesting to note here that for solution \eqref{bhole1} the field
\begin{equation} \label{singaet}
  u_\mu = \left(0,\frac{e^{C}r^{4\mu}}{1-\frac{2\kappa}{r^{2\mu+1}}}, 0,0 \right)
\end{equation}
diverges on the Killing horizon $r_h=\left(2\kappa\right)^{\frac{1}{2\mu+1}}$. This problem can be eliminated from the aether through an appropriate coordinate transformation. By performing a reparametrization of the radial variable $R$ in \eqref{bhole1b} such that
\begin{equation} \label{Rrepar}
  N(R)dR=d\rho \Rightarrow \frac{e^{\frac{C}{2}}}{\sqrt{|1-\frac{2\kappa}{(1+2\mu)R(\rho)}|}}\frac{dR}{d\rho}=1,
\end{equation}
both $u^\mu$ and $u_\mu$ assume vector components $(0,1,0,0)$ and thus are everywhere regular. Unfortunately, the explicit form of $R(\rho)$ that solves \eqref{Rrepar} cannot be written in terms of elementary functions to present it explicitly here. It is not common to have fields which diverge on the horizon, but it is not unprecedented. In \cite{Bekenstein} a black hole solution is presented with a diverging scalar field in the horizon and it is explained how this cannot produce physical effects in the geodesic motion of a particle in that particular case. In that spirit we can also argue that expression \eqref{singaet} does not really create a problem. In order to couple a relativistic particle with a vector field in a parametrization invariant manner, i.e. a Lagrangian which is homogeneous of degree one in the velocities, we would need to either multiply the square root Lagrangian $L_{rel}=\sqrt{-g_{\mu\nu}\dot{x}^\mu \dot{x}^\nu}$ with scalars of $u_\mu$, which are regular on the horizon (e.g. $u^\mu u_\mu=1$) or by adding a term similar to that of coupling an electromagnetic vector field, i.e. $\dot{x}^\mu u_\mu$, which however does not affect the dynamics because in this case it is a total derivative: $\dot{x}^\mu u_\mu=N(r)\dot{r}=\dot{G}$, where $G=\int N(r) dr$ (or alternatively because the corresponding field strength, $F_{\mu\nu}=\nabla_\nu u_\mu - \nabla_\mu u_\nu$, is zero for our choice of $u_\mu$).

At this point we need to make an important discussion. The line elements we introduced satisfy some minimal requirements that a black hole solution requires. However we need to remember that, in Lorentz violating theories, there exists the possibility of superluminal and in some cases even instantaneous motion, which makes the actual definition of a black hole rather challenging \cite{BH1,BH1b}. Even though the Killing horizon forms a trapping surface for ordinary matter it has been shown that, in Lorentz violating theories, it is possible to have wave modes with velocities faster than light. As a result information seems in principle capable of escaping outside $r=r_h$. For that matter, in theories where a preferred space-like foliation exists, the concept of the universal horizon was discovered \cite{BH2}. The latter serves as an ultimate trapping surface which can trap modes of any speed, even of infinite. In Einstein-aether theory the situation is a little better in that respect since the relevant speeds are finite \cite{BH3} and the casual structure remains similar to that of General Relativity; thus one can consider multiple horizons formed by each mode.

In our case however, we need to recognize that we consider a theory where the aether is space-like and not time-like, i.e. we do not have a theory with a preferred space-like foliation. In our case the Lorentz symmetry is violated indirectly through breaking the homogeneity of the three space with respect to the Euclidean translations. To see this consider the zero ``mass'' limit, $\kappa=0$, of \eqref{bhole1b} which yields
\begin{equation} \label{asympt}
  ds_{\kappa=0}^2 = -dt^2 + \frac{e^{-\frac{C}{2 \mu }} \left[\ln \left(\tilde{R}^{2 \mu }\right)\right]^{\frac{1}{\mu}}}{\tilde{R}^2}  \left( d\tilde{R}^2+  \tilde{R}^2 d\theta^{2}+ \tilde{R}^2 \sin^{2}\theta d\varphi^{2}\right),
\end{equation}
where we performed a transformation $R=\frac{1}{2 \mu +1}\left(2 e^{-\frac{c}{2}} \mu  \ln (\tilde{R})\right)^{\frac{2 \mu +1}{2 \mu }}$  assuming $\mu \neq 0$. Inside the parenthesis of \eqref{asympt} we see the line element of the flat three dimensional space. Due to $\mu$ not being zero however, this is multiplied by a factor that is still invariant under rotations, but not under translations. As a result the four dimensional line element is not that of the flat space and naturally it is not invariant under Lorentz transformations.

We do not know how exactly the choice of a space-like aether field we make here might affect the theory in what regards a possible superluminal motion. What is more, when the velocity of the various propagating modes is calculated in the typical Einstein-aether theory, excitations around a flat space are considered \cite{BH4}. As we mentioned however, in our configuration the flat space does not form a solution to the theory - not unless you completely remove the aether by taking  $\mu=0$, $C=0$. What is more, the aether is dynamically coupled to the scalar field through the dependence of the $\beta_i(\phi)$, which adds an additional complication. Such a calculation is highly non-trivial and to our knowledge there is no work that offers a result on this configuration. Nevertheless we want to see if the remaining freedom in the theory - due to unspecified parameters we have - can offer some way out in case there are wave modes that tend to ``break'' the Killing horizon of the space-time.

Let us first write the coupling coefficients $\beta_i(\phi)$, which are expected to enter the expressions for the velocities of the modes of the theory. For the particular solution given by \eqref{bhole1} and \eqref{solphibh1}  we obtain:
\begin{equation} \label{betas}
  \beta_1 = \frac{1}{2} \left(\frac{r}{r_h}\right)^{4 \mu } -\frac{1}{2} - \beta_3, \quad  \beta_2 = \frac{1}{2} - \frac{1}{2} \left(\frac{r}{r_h}\right)^{4 \mu },
\end{equation}
where the $\beta_3(\phi)$ remains arbitrary and there is no $\beta_4(\phi)$ since it does not affect the dynamics for our choice of aether field. In the above relations we chose to subsitute the constant $C$ appearing in \eqref{bhole1} and \eqref{solphibh1} as
\begin{equation}\label{Cchoice}
  C = \ln\left(\frac{1+2\mu}{r_h^{4\mu}}\right) .
\end{equation}
In other words we remove the hair by associating $C$ with $\mu$ and $r_h=\left(2\kappa\right)^{\frac{1}{2\mu+1}}$, the distance of the Killing horizon. We make the above choice for the following reason: In \eqref{betas} we observe that the couplings depend on the distance. If this dependence survives inside the velocities it would be crucial to have such a ratio appearing in the expressions because we could manipulate the freedom of the rest of the parameters in order to at least achieve velocities that are subluminal inside the Killing horizon $r\leq r_h$, thus the relative modes would not be able to escape it. Due to the fact that $\mu=0$ recovers the General Relativity solutions and in particular the Schwarzschild metric it is reasonable to assume that at the limit $\mu\rightarrow 0$, any velocity function $v_i^2=v_i^2\left((r/r_h)^{ \mu }\right)$ of a mode, denoted by the index $i$, that propagates has the form
\begin{equation} \label{velassumption}
  v_i^2 = 1 + h(\beta_3)\mu   \ln \left(\frac{r}{r_h}\right) +\mathcal{O}(\mu^2).
\end{equation}
In other words, a propagating mode, when $\mu=0$, must travel at most with the speed of light which in our units is $c=1$. Whether $v_i$ is to be subluminal inside the horizon $r<r_h$ is now a matter of the relative sign of $\mu$ and the function $h(\beta_3)$ which generally is to depend on the free coupling parameter that we have. Note that $r\rightarrow 0$ does not make the velocity infinite, equation \eqref{velassumption} is an approximate expression in the $\mu\rightarrow 0$ limit and the latter goes faster to zero than $\ln\left(\frac{r}{r_h}\right)$ goes to $-\infty$.

Just as a demonstration of the above, let us make a naive substitution into the velocities of the wave modes as they are derived for the typical Einstein-aether theory. As we stated above, of course we do not expect the following relations to be applicable in our case since they are derived under completely different assumptions (time-like aether, flat background, no scalar field). We just want to use them as a form of application of our argument about equation \eqref{velassumption}. The expressions of these velocities can be found in the seminal work by T. Jacobson and D. Mattingly \cite{BH4} where they use $c_i$ for the parameters of the theory instead of the $\beta_i$ we utilize here. We generally follow the conventions by Carroll et. al. \cite{BH5}, according to which the $\beta_i$ are connected to the $c_i$ of \cite{BH4} through the relations $\beta_i=\frac{c_i}{16\pi G m^2}$, where $m^2$ is the normalization constant for the aether. In a crude attempt to introduce the fact that we use a space-like aether we set $m^2=-1$ in these relations\footnote{This part is not crucial in the analysis, similar results can be obtained for $m^2=1$, but for a different choice for the signs of the parameters.}. At the same time, according to the action we adopted \eqref{ac.01}, we have in our units $8\pi G =1$. Thus, we use the identification $\beta_i=\frac{-c_i}{2}$ and write the relations obtained in \cite{BH4} as
\begin{align}
  v_2^2 & = \frac{1}{2 \beta_1+2 \beta_3+1}\\
  v_1^2 & = \frac{\beta_1^2+\beta_1-\beta_3^2}{(2 \beta_1+2 \beta_3+1) (\beta_1+\beta_4)} \\
  v_0^2 & = -\frac{(\beta_1+\beta_4+1) (\beta_1+\beta_2+\beta_3)}{(2 \beta_1+2 \beta_3+1) (\beta_1+\beta_4) (\beta_1+3 \beta_2+\beta_3-1)} .
\end{align}
The mode with velocity $v_0$ does not propagate in our case since $\beta_1+\beta_2+\beta_3=0$, see \eqref{betas}. By using the latter in the remaining two expressions we have
\begin{align}
  v_2^2 & = \left(\frac{r}{r_h}\right)^{-4\mu} \simeq 1 - 4\mu \ln \left(\frac{r}{r_h}\right) + \mathcal{O}(\mu^2)\\
  v_1^2 & = \frac{4 \beta_3 h^{4 \mu }+h^{8 \mu } r^{-4 \mu }-r^{4 \mu }}{(4 \beta_3+2) h^{4 \mu }-2 r^{4 \mu }} \simeq 1 + \frac{4 \mu ^2 }{\beta_3}\left(\ln \left(\frac{r_h}{r}\right)\right)^2 + \mathcal{O}(\mu^3) ,
\end{align}
where we observe what we wanted to demonstrate with \eqref{velassumption}. In order to keep $v_2$ subluminal inside the horizon $r\leq r_h$, where $\ln \left(\frac{r}{r_h}\right)\leq 0$, we need to consider $\mu<0$. At the same time we see that as $\mu\rightarrow 0^-$ the velocity $v_1$ can be kept smaller that the speed of light by simply requiring $\beta_3<0$. In fact $\beta_3$ can also depend on $\phi$ and through this on the ratio $\frac{r}{r_h}$. As a result we can just alternatively choose $\beta_3$ to make zero $v_1$ altogether. However we see that even under the more restrictive assumption $\beta_3=$constant we can still manipulate the expressions successfully. 

Of course from the moment we do not have the exact relations for the particular theory we consider, we cannot be certain that the propagating modes can truly be controlled in this manner. However, the fact that the solution is continuously and smoothly connected to the one from General Relativity as $\mu\rightarrow 0$, together with the remaining freedom we have in some of the parameters, indicates that there is hope that under certain restrictions the black hole definition does not completely break down. What is more, the fact that the black hole hair might be important in this respect, through their removal by associating them with the rest of the parameters, is particularly interesting.

\subsection{The case of cosmological constant}

It is easy to add a cosmological constant $\Lambda$ to the previous solution
(\ref{bhole1}). In particular, we observe that the line element
\begin{equation}
ds^{2}=-\left(  1-\frac{2\kappa}{r^{2\mu+1}}+lr^{2}\right)
dt^{2}+\frac{e^{C}r^{4\mu}}{1-\frac{2\kappa}{r^{2\mu+1}%
}+lr^{2}}dr^{2}+r^{2}\left(  d\theta^{2}+\sin^{2}\theta d\varphi^{2}\right),
\end{equation}
where $l$ is a constant, together with the same expression for the scalar field given by \eqref{solphibh1},
satisfy the field equations with $F(\phi)=\mu\phi^{2}$, $M(\phi)=\nu\phi^{2}$,
$V(\phi)=\Lambda$, with $l=-\frac{2\mu+1}{2\mu+3}\Lambda$. That is it forms the solution when a cosmological constant $\Lambda$ is also considered.

However we must note that another singularity is added in this case when $r\rightarrow+\infty$ for $-\frac{1}%
{2}<\mu<0$ since the scalar curvature is now
\begin{equation}
\mathcal{R}=\frac{2}{r^{2}}+2e^{-C}\left(  \frac{8(\mu-1)\mu\kappa%
}{r^{3(2\mu+1)}}+\frac{(4\mu-1)}{r^{2(2\mu+1)}}+\frac{6l(\mu-1)}{r^{4\mu}%
}\right)  .
\end{equation}
This means that in the presence of a cosmological constant we have to further
restrict $\mu$ to be positive so that the second singularity at infinity is avoided. Again, we notice that at the limit $\mu\rightarrow 0$, $C=0$, the known from General Relativity solution with a cosmological constant is obtained with $\mathcal{R}=-12 l$.

\subsection{The case of electrostatic field}

Alternatively (or in addition to the above) one may consider an appropriate
electrostatic field. In this case we obtain
\begin{equation}%
\begin{split}
ds^{2}= &  -\left(  1-\frac{2\kappa}{r^{2\mu+1}}+\left(
1-4\mu^{2}\right)  \frac{e^{-C}Q^{2}}{r^{2}}\right)  dt^{2}\\
&  +\frac{e^{C}r^{4\mu}}{1-\frac{2\kappa}{r^{2\mu+1}%
}+\left(  1-4\mu^{2}\right)  \frac{e^{-C}Q^{2}}{r^{2}}}dr^{2}+r^{2}\left(
d\theta^{2}+\sin^{2}\theta d\varphi^{2}\right)
\end{split}
\end{equation}
which corresponds to a solution in the presence of an electric field with
potential $U(r)=\frac{Q}{r^{1-2\mu}}$, if we include in the right hand side of \eqref{EinA} the energy momentum tensor
\begin{equation}\label{enmomem}
  T^{EM}_{\mu\nu} = 2F_{\mu\kappa}F_\nu^{\; \kappa} - \frac{1}{2} g_{\mu\nu} F_{\kappa\lambda}F^{\kappa\lambda},
\end{equation}
where $F_{\mu\nu} = \nabla_\mu A_\nu - \nabla_\nu A_\mu$ and $A = U(r) dt$. Once more the solution is compatible with the expression \eqref{solphibh1} for the scalar field.

\section{Existence of circular orbits}

\label{sec4}

In this section we investigate whether the new black hole solution \eqref{bhole1} supports stable trajectories of massive particles.

Lets assume the affinely parametrized geodesic equations for a time-like particle in a space-time whose line element is
(\ref{bhole1}). The system is described by the Lagrangian
\begin{equation}
L_{Geodesic}=-\frac{1}{2}\zeta\left(  r\right)  \left(  \frac{dt}{ds}\right)
^{2}+\frac{1}{2}\frac{e^{C}r^{4\mu}}{\zeta\left(  r\right)  }\left(  \frac
{dr}{ds}\right)  ^{2}+\frac{1}{2}r^{2}\left(  \frac{d\Omega}{ds}\right)
^{2},\label{p01}%
\end{equation}
where $d\Omega^{2}= d\theta^{2}+\sin^{2}\theta d\varphi^{2}$
and $\zeta\left(  r\right)  =1-\frac{2\kappa}{r^{2\mu+1}}$.
The geodesic equations admit the conservation laws%
\begin{equation}
~\zeta\left(  r\right) \frac{dt}{ds}  =\mathcal{E} \quad
, \quad r^{2}\sin^2\theta \frac{d\varphi}{ds}  =L_0~~
\end{equation}
while the Hamiltonian of (\ref{p01}) in the equatorial plane $\theta=\frac{\pi}{2}$ and for the case of a timelike particle yields
\begin{equation}
e^{C}r^{4\mu}\left(  \frac{dr}{ds}\right)  ^{2}=  \mathcal{E}%
^{2}-\left(  2+\frac{L_0^{2}}{r^{2}}\right)  \zeta\left(  r\right)
\end{equation}
As previously, we define the new variable $dR=r^{2\mu}dr$ which gives $R=\frac{1}{2\mu
+1}r^{2\mu+1},$ hence the latter equation becomes%
\begin{equation}
e^{C}\left(  \frac{dR}{ds}\right)  ^{2}= \mathcal{E}^{2}-V_{g}\left(
R\right)
\end{equation}
where now%
\begin{equation}
V_{g}\left(  R\right)  =2\left(  1-\frac{\bar{K}}{R}\right)  +\frac{\bar
{L}^{2}}{R^{\bar{\mu}}}-~\bar{K}\frac{\bar{L}^{2}}{R^{1+\bar{\mu}}%
}\label{p.02}%
\end{equation}
and $\bar{L}^{2}=\left(  \frac{\bar{\mu}}{2}\right)  ^{\frac{\bar{\mu}}{2}}L_0^{2}%
,~\bar{K}=\bar{\mu}\kappa$ and $\bar{\mu}=\frac{2}{1+2\mu}$.
For real valued scalar field, that is, for $-\frac{1}{2}<\mu<0$, i.e.
$\bar{\mu}>0$ and for large values of $R$, which also means large values of
$r$,  the dominant terms are those of the Newtonian potential, that is
$V_{g}\left(  R\right)  \simeq2\left(  1-\frac{\bar{K}}{R}\right)  $.

\begin{figure}[ptb]
\centering\includegraphics[width=0.5\textwidth]{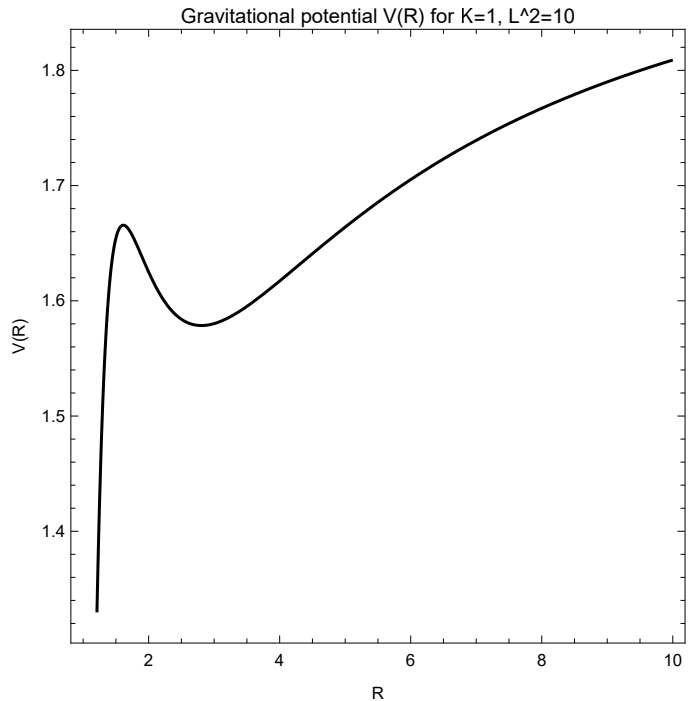} \caption{Evolution
of the gravitational potential $V_{g}\left(  R\right)  =2\left(  1-\frac{1}%
{R}\right)  +\frac{10}{R^{3}}-~\frac{10}{R^{4}}.$ }%
\label{fig0}%
\end{figure}

We consider the special case where $\bar{K}=1,~\bar{L}^{2}=10$ and $\bar{\mu
}=3$, then  $V_{g}\left(  R\right)  =2\left(  1-\frac{1}{R}\right)  +\frac
{10}{R^{3}}-~\frac{10}{R^{4}}$ and $\frac{d}{dR}\left(  V_{g}\left(  R\right)
\right)  =\frac{2}{R^{5}}\left(  20-15R+R^{3}\right)  $, thus the stationary
point $\frac{d}{dR}\left(  V_{g}\left(  R\right)  \right)  =0$ is found on the
positions $R_{1}\simeq1.61$ and $R_{2}\simeq2.81$. For this set of values the Killing
horizon is located on $R=1$, thus both stationary points are outside of the
latter. Point $R_{1}$ is unstable, that is $\frac{d^{2}}{dR^{2}}\left(
V_{g}\left(  R\right)  \right)  |_{R\rightarrow R_{1}}<0$ while point $R_{2}$ is an
attractor, since $\frac{d^{2}}{dR^{2}}\left(  V_{g}\left(  R\right)  \right)
|_{R\rightarrow R_{2}}>0.$ In Fig. \ref{fig0} we present potential $V_{g}\left(
R\right)  $ for $\bar{K}=1,~\bar{L}^{2}=10$ from where it is clear that there
are periodic solutions around the stationary point $R_{2}$

\begin{figure}[ptb]
\centering\includegraphics[width=0.5\textwidth]{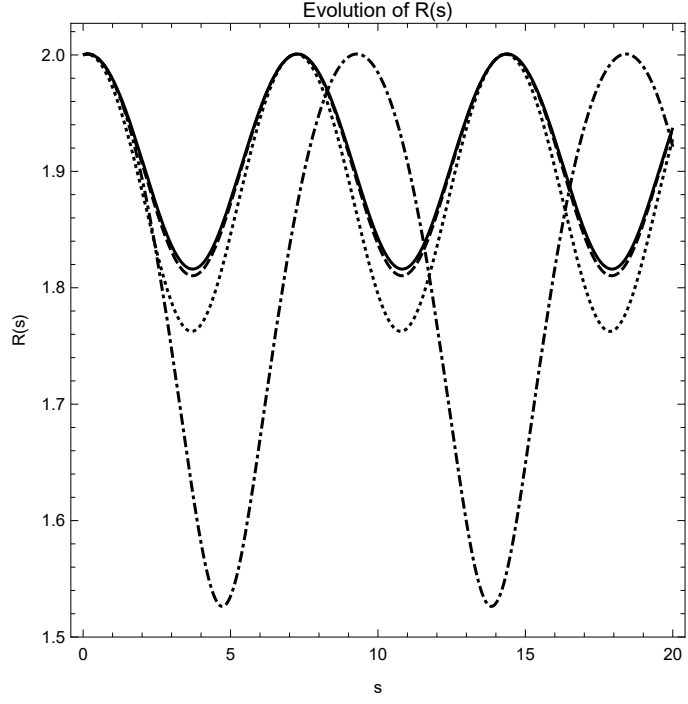}
\caption{Qualitative evolution of $R\left(  s\right)  $ around the periodic
solution for various values of $\bar{\mu}$. Solid line is for $\bar{\mu}=2,$
which correspond to the Schwarzschild solution ($\mu=0$), dashed line is for
$\bar{\mu}=2.01$, dotted line is for $\bar{\mu}=2.1$ and dashed-dotted line is
for $\bar{\mu}=3$. The evolution is for  $\bar{K}=1$,~$\bar{L}^{2}=3$ and
$C=0.$}%
\label{fig1}%
\end{figure}

In order to understand the deviation from the Schwarzschild solution we
consider $\bar{K}=1$,~$\bar{L}^{2}=3$ and in Fig. \ref{fig1} we present the
qualitative evolution of $R\left(  s\right)  $ for various values of $\bar
{\mu}$. We observe that there is a deviation from the Schwarzschild solution
in the period of the oscillation. The trajectories presented in Fig.
\ref{fig1} are for the same initial conditions.

\section{Conclusions}

\label{sec5}

In this work we studied the existence of exact static spherically symmetric
solutions in Einstein-Aether scalar field gravity with an interaction
between the scalar field and the Aether. For the gravitational theory proposed
by Kanno and Soda \cite{Kanno:2006ty} and for the requirement the field equations to admit a
point-like Lagrangian we were able to find analytic and exact solutions for
specific functional forms of the unknown functions. We distinguished conditions that are necessary for certain black hole space-times to emerge. The solutions correspond to a massless scalar field and can be modified appropriately to introduce a cosmological constant and/or an electrostatic field to contribute in the effective fluid.

The black hole solutions that we presented admit an additional (apart from the mass) constant of
integration $C$ in the metric. The latter emanates from the scalar content of the theory and allows as to characterize these solutions as hairy black holes. Surprisingly enough, a limit exists that connects the resulting space-time to the known static black hole solutions of General Relativity.  When we have $\mu\rightarrow0$, $C\rightarrow 0$, the solutions that we found tend to the
Schwarzschild, the de-Sitter Schwarzschild and Reissner-Nordstr\"{o}m metrics.
Hence, we can say that $\mu$ in these cases appears as a measure of the
radial modification produced due to the aether having a velocity in the $r$
direction, that is, due to the Lorentz violation.

We need to mention that the solutions we derived so far correspond to specific combinations of the
free parameters  $\mu$ and $\nu$. However, these combinations are not unique in giving black hole
space-times. For instance, if we take $\mu=-\frac{1}{6}$, $\nu=1$ we get the exact
solution
\begin{equation}
ds^{2}=-r^{\frac{16}{5}}\left(  1-\frac{2\kappa}{r^{2}}\right)  ^{3}%
dt^{2}+\frac{e^{C}r^{-\frac{6}{5}}}{\left(  1-\frac{2\kappa}{r^{2}}\right)
^{3}}dr^{2}+r^{2}\left(  d\theta^{2}+\sin^{2}\theta d\varphi^{2}\right)
\end{equation}
with scalar field
\begin{equation}
\phi(r)=\pm\sqrt{\frac{5}{3}}\frac{e^{\frac{C}{2}}}{r^{\frac{3}{5}}\left(
1-\frac{2\kappa}{r^{2}}\right)  },
\end{equation}
which is distinct from what we had previously. The resulting space-time admits a curvature singularity at the origin $r=0$, while $r=\sqrt{2\kappa}$ is a coordinate singularity. This indicates that there is a rich structure in the theory which leads to various black hole solutions.

Finally, it is straightforward to see that the above results have their cosmological
counterparts which are obtained through the transformations: $t\leftrightarrow
r,~\phi\mapsto i\phi$. That is, we need only interchange $t$ with $r$ in the
above line elements and wherever we have $\phi$, put in its place $i\phi$. For
example if we take \eqref{bhole1} and \eqref{solphibh1} the cosmological dual
solution is
\begin{equation}
ds^{2}=-\frac{e^{C}t^{4\mu}}{1-\frac{2\kappa}{t^{2\mu+1}}%
}dt^{2}+\left(  1-\frac{2\kappa}{t^{2\mu+1}}\right)
dr^{2}+t^{2}\left(  d\theta^{2}+\sin^{2}\theta d\varphi^{2}\right)
\end{equation}
with
\begin{equation}
\phi(t)=\pm\frac{e^{\frac{C}{2}}t^{2\mu}}{\sqrt{\mu(2\mu+1)}},
\end{equation}
and it corresponds to a theory with $F=-\mu\phi^{2}$ and $M=-\nu\phi^{2}$.
Since there is no obligation for having an horizon here, $\kappa$ can be taken negative. We remark that in the case of
cosmological solutions the aether field is timelike.

In a future work we plan to extend our study on the thermodynamics properties
of these new black hole solutions.

\begin{acknowledgments}
AP\ \& GL were funded by Agencia Nacional de Investigaci\'{o}n y Desarrollo -
ANID through the program FONDECYT Iniciaci\'{o}n grant no. 11180126.
Additionally, GL is supported by Vicerrector\'{\i}a de Investigaci\'{o}n y
Desarrollo Tecnol\'{o}gico at Universidad Catolica del Norte.
\end{acknowledgments}


\begin{thebibliography}{99}                                                                                               %


\bibitem {jacob}T.~Jacobson and D.~Mattingly, Phys.\ Rev.\ D \textbf{64}, 024028 (2001)

\bibitem {DJ2}W.~Donnelly and T.~Jacobson, Phys.\ Rev.\ D \textbf{82}, 081501 (2010)

\bibitem {Carru}I.~Carruthers and T.~Jacobson, Phys.\ Rev.\ D \textbf{83}, 024034 (2011)

\bibitem {carroll}S.~M.~Carroll and E.~A.~Lim, Phys.\ Rev.\ D \textbf{70}, 123525 (2004)

\bibitem {mm1}C. Eling, T. Jacobson and D. Mattingly, Deserfest: A celebration
of the life and works of Stanley Deser. Proceedings, Meeting, Ann Arbor, USA,
April 3-5, (2004) [gr-qc/0410001]

\bibitem {hh1}H. Wei, X.-P. Yan and Y.-N. Zhou, Gen. Rel. Grav. \textbf{46}, 1719 (2014)

\bibitem {hh2}R. Chan, M.F.A. de Silva and V.H. Satheeshkumar, JCAP \textbf{05}, 025 (2020)

\bibitem {hh3}A.\ Casalino, L. Sebastiani and S. Zerbini, Phys. Rev. D \textbf{101},
104059 (2020)

\bibitem {ea1}C. Heinicke, P. Baekler and F.W. Hehl, Phys. Rev. D \textbf{72}, 025012 (2005)

\bibitem {Horava:2009uw}P.~Horava, Phys.\ Rev.\ D \textbf{79}, 084008 (2009)

\bibitem {esf}D. Garfinkle and T. Jacobson, Phys. Rev. Lett. \textbf{107}, 191102 (2011)

\bibitem {esf1}T. Jacobson, Phys. Rev. D \textbf{89}, 081501 (2014)

\bibitem {TJab13}T. Jacobson, Phys.\ Rev.\ D \textbf{89}, 081501 (2014).

\bibitem {sot01}T.P. Sotiriou, J. Phys. Conf. Ser. \textbf{283}, 012034 (2011)

\bibitem {DJ}W.~Donnelly and T.~Jacobson, Phys.\ Rev.\ D \textbf{82}, 064032 (2010)

\bibitem {Kanno:2006ty}S.~Kanno and J.~Soda, Phys.\ Rev.\ D \textbf{74}, 063505 (2006)

\bibitem {ea2}X. Meng and X. Du, Phys. Lett. B \textbf{710}, 493 (2012)

\bibitem {Barrow:2012qy}J.~D.~Barrow, Phys.\ Rev.\ D \textbf{85}, 047503 (2012)

\bibitem {Carruthers:2010ii}I.~Carruthers and T.~Jacobson, Phys.\ Rev.\ D \textbf{83},
024034 (2011)

\bibitem {Latta:2016jix}J.~Latta, G.~Leon and A.~Paliathanasis, JCAP \textbf{1611}, 051
(2016) \

\bibitem {Alhulaimi:2017ocb}B.~Alhulaimi, R.~J.~Van Den Hoogen and
A.~A.~Coley, JCAP \textbf{1712}, 045 (2017)

\bibitem {VanDenHoogen:2018anx}R.~J.~Van Den Hoogen, A.~A.~Coley,
B.~Alhulaimi, S.~Mohandas, E.~Knighton and S.~O'Neil, JCAP \textbf{1811}, 017 (2018)

\bibitem {Roumeliotis:2019tvu}M.~Roumeliotis, A.~Paliathanasis, P.~A.~Terzis
and T.~Christodoulakis, arXiv:1911.03660 [gr-qc]

\bibitem {Roumeliotis:2018ook}M.~Roumeliotis, A.~Paliathanasis, P.~A.~Terzis
and T.~Christodoulakis, Eur.\ Phys.\ J.\ C \textbf{79}, no. 4, 349 (2019)

\bibitem {Paliathanasis:2020bgs}A.~Paliathanasis, Phys. Rev. D \textbf{101}, 064008 (2020)

\bibitem {Paliathanasis:2019pcl}A.~Paliathanasis, G.~Papagiannopoulos,
S.~Basilakos and J.~D.~Barrow, Eur.\ Phys.\ J.\ C \textbf{79}, no. 8, 723 (2019)

\bibitem {palgn1}A. Paliathanasis and G.\ Leon, Eur. Phys. J. C \textbf{80}, 355 (2020)

\bibitem {palgn2}A. Paliathanasis and G. Leon, Eur. Phys. J. C \textbf{80}, 589 (2020)

\bibitem {Coley:2015qqa}A.~A.~Coley, G.~Leon, P.~Sandin and J.~Latta, JCAP \textbf{1512}, 010 (2015)

\bibitem {Coley:2019tyx}A.~Coley and G.~Leon, Gen.\ Rel.\ Grav.\ \textbf{51}, no. 9,
115 (2019)

\bibitem {Leon:2019jnu}G.~Leon, A.~Coley and A.~Paliathanasis, Annals
Phys.\ 412, 168002 (2020)

\bibitem {bhea1}C. Eling and T. Jacobson, Class. Quantum Grav. \textbf{23}, 5643 (2010)

\bibitem {bhea2}T.\ Tamaki and U. Miyamoto, Phys. Rev. D \textbf{77}, 024026 (2008)

\bibitem {bhea3}R.A. Konoplya and A. Zhidenko, Phys. Lett. B \textbf{648}, 236 (2007)

\bibitem {bhea4}K. Lin, F.-H. Ho and W.-L. Qian, IJMPD \textbf{28}, 1950049 (2019)

\bibitem {bhea5}R.A. Konoplya and A. Zhidenko, Phys. lett. B \textbf{644}, 186 (2007)

\bibitem {bhea6}C. Ding, Nucl. Phys. B \textbf{938}, 736 (2019)

\bibitem {bhea7}M. Umair Shahzad, R. Ali, A. Jawad and S. Rani, Chin. Phys. C
\textbf{44}, 065106 (2020)

\bibitem {bhea8}C. Ding, C. Liu, R. Casana and A. Cavalcante, Eur. Phys. J. C
\textbf{80}, 178 (2020)

\bibitem {bh01}E. Barausse, T.P. Sotiriou and I. Vega, Phys.\ Rev.\ D \textbf{93},
044044 (2016)

\bibitem {bh02}C. Eling, T. Jacobson and M.C. Miller, Phys. Rev.\ D \textbf{76}, 042003
(2007); Erratum: Phys. Rev. D 80, 129906 (2009)

\bibitem {bh03}M. Bhattacharjee, S. Mukohyama, M.-B. Wan and A. Wang,
Phys.\ Rev. D \textbf{98}, 064010 (2018)

\bibitem {bh04}G. Panotopoulos, D. Vernieri and I. Lopes, Eur. J. Phys. C \textbf{80},
537 (2020)

\bibitem {bh05}C. Ding, Phys. Rev. D \textbf{96}, 104021 (2017)

\bibitem {bh06}D. Garfinkle, C. Eling and T. Jacobson, Phys.\ Rev. D \textbf{76},
024003 (2007)

\bibitem{Carrollsp} L. Ackerman, S. M. Carroll and M. B. Wise Phys. Rev. D \textbf{75} 083502 (2007)

\bibitem{Dulaney} T. R. Dulaney, M. I. Gresham and M. B. Wise Phys.Rev.D \textbf{77} 083510 (2008); Phys. Rev. D \textbf{79}, 029903(E) (2009)

\bibitem {nn5}T.G. Rizzo, JHEP 09, 036 (2005)

\bibitem {nn4}S. M. Carroll and H. Tam, Phys. Rev. D \textbf{78}, 044047 (2008)

\bibitem {nn3}A. Chatrabhuti, P. Patcharamaneepakorn, and P. Wongjun, JHEP 08,
019 (2009)

\bibitem {nn1}M. Gurses and C. Senturk, Commun. Theor. Phys. \textbf{71}, 312 (2019)

\bibitem {nn2}A. \"{O}vg\"{u}n, \.{I}. Sakalli and J. Saavedra, to appear in
Chinese Physics C [DOI: 10.1088/1674-1137/abb532]

\bibitem{unstable1} B. Himmetoglu, C. R. Contaldi and M. Peloso, Phys. Rev. Lett. \textbf{102} 111301 (2009)

\bibitem{stable1} C. Armendariz-Picon and  A. Diez-Tejedor, JCAP \textbf{12} 018 (2009)

\bibitem{GenlyAndrNik} G. Leon, A. Paliathanasis and N. Dimakis, arXiv:2010.02775 [gr-qc]

\bibitem{Papadopoulos} G. O. Papadopoulos J. Math. Phys. \textbf{47}, 092502 (2006)

\bibitem{EinMaxdil} K. C. K. Chan, J. H. Horne and R. B. Mann, Nucl. Phys. B \textbf{447}, 441 (1995)

\bibitem{Bekenstein} J. D. Bekenstein, Annals Phys. \textbf{91}, 75 (1985)

\bibitem{BH1} E. Barausse, T. Jacobson and T. P. Sotiriou, Phys. Rev. D \textbf{84}, 124043 (2011)

\bibitem{BH1b} E. Barausse and T. P. Sotiriou, Class. Quantum Grav. \textbf{30}, 244010 (2013)

\bibitem{BH2} J. Bhattacharyya, M. Colombo and T. P. Sotiriou, Class. Quantum Grav. \textbf{33},  235003 (2016)

\bibitem{BH3} D. Blas and S. Sibiryakov, Phys. Rev. D \textbf{84}, 124043 (2011)

\bibitem{BH4} T. Jacobson and D. Mattingly, Phys. Rev. D \textbf{70},  024003 (2004)

\bibitem{BH5} S. M. Carroll, T. R. Dulaney, M. I. Gresham and H. Tam, Phys. Rev. D \textbf{79}, 065011 (2009)

\end{thebibliography}
\end{document}